\documentclass[reprint,amsmath,amssymb,aip,pre,author-numerical]{revtex4-1}
\usepackage{graphics}
\usepackage{graphicx}
\usepackage{dcolumn}% Align table columns on decimal point
\usepackage{bm}
\usepackage{natbib}
\usepackage{hyperref}
\usepackage{color}
\usepackage{physics}
\newcommand*{\ie}{i.e., }
\usepackage{tikz}
\usepackage{siunitx}
\usepackage[utf8]{inputenc}

\usepackage{todonotes}

\begin{document}
\title[pH-gradient-driven membrane deformations]{A model of membrane deformations driven by a surface pH gradient}
\author{Toni V. Mendes}
\affiliation{Laboratoire Ondes et Mati\`ere d'Aquitaine, Universit{\'e} de Bordeaux, Unit\'e Mixte de Recherche 5798, CNRS, F-33400 Talence, France}
\affiliation{Sorbonne Universit{\'e}, CNRS, Laboratoire de Physique Th{\'e}orique de la Mati{\`e}re Condens{\'e}e (LPTMC, UMR 7600), F-75005 Paris, France}
\author{Jonas Ranft}
\affiliation{Institut de Biologie de l'ENS, Ecole Normale Sup{\'e}rieure, CNRS, Inserm, Université PSL, 46 rue d'Ulm, F-75005 Paris, France}
\author{H\'el\`ene Berthoumieux}
\affiliation{Sorbonne Universit{\'e}, CNRS, Laboratoire de Physique Th{\'e}orique de la Mati{\`e}re Condens{\'e}e (LPTMC, UMR 7600), F-75005 Paris, France}
\affiliation{Fachbereich Physik, Freie Universität Berlin, Arnimallee 14, Berlin,14195, Germany}

\begin{abstract}
	Many cellular organelles are membrane-bound structures with complex membrane composition and shape. Their shapes have been observed to depend on the metabolic state of the organelle, and the mechanisms that couple biochemical pathways and membrane shape are still actively investigated. Here, we study a model coupling 
	inhomogeneities in the lipid composition and membrane geometry
	%in 
	via %
	a generalized Helfrich free energy. We derive the resulting stress tensor, the Green's function for a tubular membrane and compute 
	the phase diagram of the induced deformations. We then apply this model to study the deformation of mitochondria cristae described as
     membrane tubes supporting a pH gradient 
	at its surface %
	This gradient in turn controls the 
	lipid %
	composition of the membrane via the protonation/deprotonation of cardiolipins, which are acid-based lipids known 
	to be crucial for mitochondria shape and functioning. Our model predicts the appearance of tube 
	%radius modulations 
	deformations %
	resembling  the observed shape changes of cristea when submitted to a 
	%high 
	proton gradient.
\end{abstract}

\maketitle
\section{Introduction}
Lipid membranes are fundamental components of cells as they compartmentalize space. Notably, many organelles important for cellular function such as the endoplasmatic reticulum, the Golgi apparatus, or mitochondria are membrane-enclosed structures. For this last case, the lipid membrane also takes part in biochemical processes that it encloses by facilitating the 2D diffusion of reactants (protons) ~{\cite{branden2006}}. 
Physical descriptions of  membranes as two-dimensional 
sheets with a mechanical energy given by the Helfrich model~\cite{helfrich1973}, which in addition to surface tension assumes a quadratic dependence of the energy on the intrinsic mean curvature of the 
surface, have been very successful in capturing the rich variety of shapes observed for 
\emph{in vitro} systems~\cite{seifert1997}.
Since Helfrich's seminal work, many extensions to his model were proposed to take into account internal degrees of freedom of the membrane such as variations of membrane mass density and/or composition by way of introducing additional 
terms in the Hamiltonian\cite{seifert1995,AF}. Beyond scalar degrees of freedom, vectorial fields such as a local lipid tilt can be considered with this approach~\cite{lubensky92,hamm2000}.

At a first glance, Gaussian Hamiltonians for membrane mechanics are simple functionals that can be constructed using a Landau-Ginzburg expansion of the surface energy. 
However, obtaining the equilibrium shape of the surface by a straightforward minimization of the Hamiltonian can become very tricky, and sometimes proves impractical. 
The variational calculus expressed in terms of differential geometry  (characterizing the surface 
by its intrinsic basis, 
metric, %
curvature tensors etc.) becomes quickly complex and cumbersome. Illustrating these difficulties, the ``shape equation'' associated with the Helfrich Hamiltonian was derived more than ten 
years after the introduction of the Helfrich model~\cite{helfrich1989}. In an important development, Guven and coworkers made the minimization process much more elegant and easy by the introduction of a constrained functional~\cite{CG02,Capovilla02,Guv04a}. In short, the geometrical relations imposed by surface continuity are enforced using Lagrange multipliers, one of which being 
%identified as 
the 
surface %
stress tensor~\cite{Deserno_curvatureconvention}. The general identification of the stress tensor for any Helfrich-type model and 
%any 
arbitrary %
geometry now makes the exploration of resulting membrane shapes considerably simpler.

Besides lipid membranes, the cell cortex is another fundamental  cellular surface. Remarkably, a stochastic chemico-mechanical energy conversion driven by ATP hydrolysis can lead to a local increase of its mechanical tension and internal torque~\cite{berthoumieux2014,Salbreux2017}. From a more general point of view, the %
mechanics of these out-of-equilibrium surfaces can be described via constitutive relations for tensions, torques and surface chemical fluxes including 
%the 
so-called %
active terms that would not exist in passive surfaces~\cite{berthoumieux2014,Salbreux2017}. This coupling between active agent concentration and  surface mechanics gives rise to the spontaneous formation of nontrivial shapes~\cite{mietke}. Lipid vesicles can also deform or divide in response to 
chemical stimuli and the development of such minimal models for biological self-reproduction has gained attention in last decades~\cite{karimi2018}. In particular, pH variation was shown to affect the bending modulus of bilayers 
and to generate tubular 
invaginations~\cite{khalifat2008membrane}. %

Proton diffusion along biological membranes is essential for 
\emph{in cellulo} %
energy production, and the role of the membrane composition in this mechanism is a domain of active research~\cite{hugentobler2020,joubert2021}. In mitochondria, the shape 
of the inner membrane invaginations (cristae) enclosing ATP production varies with rate of ATP synthesis which itself is coupled to a flux of protons diffusing on the membrane ~\cite{mannella2006structure,cogliati2016mitochondrial}.
Inspired by this observation, a 
model of a tubular membrane described by a phenomenological pH-dependent %
Helfrich Hamiltonian and submitted to a spatially modulated %
proton flux was proposed~\cite{patil2020}.
It could  reproduce some qualitative observations of the coupling between cristae  
shape and metabolic state of the organelle. 

In this work, we study the shapes of inhomogeneous membranes 
that are composed of two lipids 
which ratio is driven by an external surface chemical flux. We 
then apply this model to decribe  
mitochondria cristae shapes as a function of 
a varying proton flux. The work is organized as follows. We start by introducing an Helfrich Hamiltonian which, in addition 
to the geometrical 
terms, includes a contribution  
related to the lipid composition and mass density based on %
a  Landau-Ginzburg %
approach. In particular we take into account 
coupling terms between the local curvature and these physical parameters. 
%Thus, we 
We then %
derive the stress tensor of this model in an intrinsic
surface %
reference frame. We consider the case of a cylindric geometry and compute the Green's function of the system, defined as the expression 
%of 
for %
the deformation field for a punctual perturbation in the membrane composition. The richness of the phase diagram suggests that such a model could be relevant to describe {\it in vivo} membranes.
We finally apply our 
formalism to model the shape of the mitochondrial membrane invaginations, the cristae, as finite tubes driven by a proton flux of 
varying %
intensity. We show that this model reproduces qualitatively the shape observed {\it in vivo} and the shape change between 'active' mitochondria (state III) and mitochondria in a rest state (state IV)~\cite{mannella2006structure}. 
We conclude with a summary and a discussion of our results. %

%\section{Shape of an inhoumogeneous membrane}
\section{Inhomogeneous membranes: Landau-Ginzburg Hamiltonian and resulting shapes} %

%\subsection{Model}
\subsection{Helfrich model with composition-dependent free energy} %

\begin{figure*}
 		\includegraphics[scale=0.9]{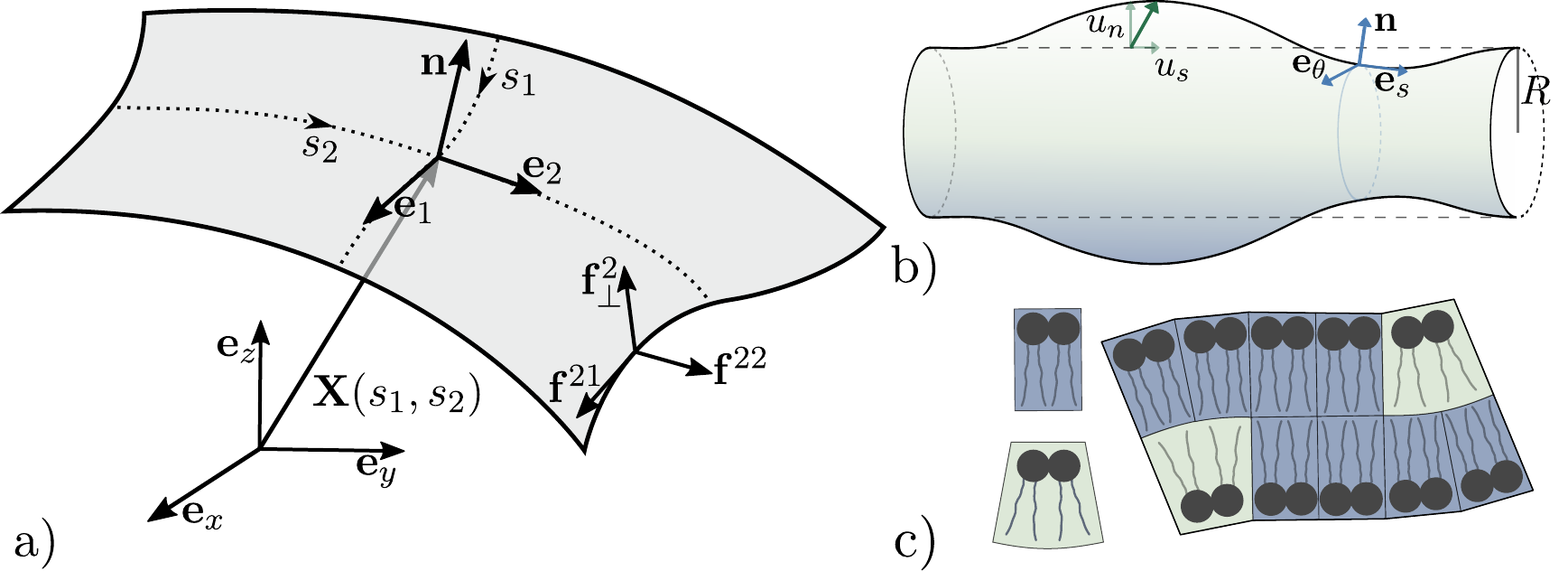} % AR : 0.6
 	\caption{(a) Representation of a 3D surface, a local basis $(\vb{e}_1$, $\vb{e}_2)$, a normal vector $\vb{n}$ and the stress tensor showing the $3$ possible stresses one can apply on a surface cut. b) Cylinder of radius $R$ deformed by the field $\vb{u}=u_n \vb{n} + u_s \vb{e}_s$. c) Sketch of a cylindric (blue) and a conic (green) lipid and of a membrane with a curvature induced by composition inhomogeneity.}
 	\label{fig:1}
 \end{figure*}

We consider
a curved surface, 
the Cartesian coordinates 
of which are given by 
the 3D vector 
field ${\bf  X} (s_1,s_2)$. The 
surface is parametrized by two generalized coordinates $s_1$, $s_2$. 
One can then obtain the local intrinsic basis of $\bf {X}$,  $({\bf e}_1, {\bf e}_2)$,  defined as ${\bf e}_a =\partial_a {\bf X}$, $(a=1,2)$, and the normal vector ${\bf n}={\bf e}_1 \times {\bf e}_2/| {\bf e}_1 \times {\bf e}_2|$, 
a vector 
of unit length 
perpendicular to 
the 
surface, see Fig.~\ref{fig:1}a.
We can furthermore define the two fundamental forms of the surface, 
i.e.~the metric tensor 
$g_{ab}={\bf e}_a \cdot {\bf e}_b$ 
and the curvature tensor $K_{ab}={\bf e}_a \cdot \partial_b {\bf n}$. The surface element $dA$  is equal to $dA=\sqrt{g} ds_1ds_2$, where $g =\det g_{ab}$ 
is the determinant of the metric tensor. 
For the tensorial calculations below, 
we remind the  Einstein summation convention and the passage from covariant to contravariant coordinates, such that $T^a_b=T_{bk}g^{ka} = \sum_{k=1,2} T_{bk}g^{ka}$, with $g^{ab}=g_{ab}^{-1}$, for 
${\bf T}$ a tensor of rank 2.

Let us now consider a membrane composed of two lipids 
$L_1$ and $L_2$ linked by a chemical reaction $L_1 \rightleftharpoons L_2$, 
with respective surface mass densities $\rho_{L1}(s_1,s_2)$ and $\rho_{L2}(s_1,s_2)$. 
We introduce two scalar fields to describe the internal degrees of freedom of this inhomogeneous membrane: the mass density  $\rho(s_1,s_2)=\rho_{L1} + \rho_{L2}$ %
of the two lipids and the relative mass fraction $\phi (s_1,s_2)= {\rho_{L2}}/{\rho}$ %
of the lipid $L_2$, which gives the local lipid composition. We then define 
a homogeneous reference state $(\phi_0,\rho_0)$  for the system. 
In this state, the 
mass density $\rho_0$ and the composition $\phi_0$ are constant along the surface. Finally, we characterize 
inhomogeneous states %
of the membrane  using
\begin{equation}
	r(s_1,s_2)=\frac{\rho(s_1,s_2)-\rho_0}{\rho_0}, \quad \psi(s_1,s_2)=\frac{\phi(s_1,s_2)-\phi_0}{\phi_0}
\end{equation}
where $r(s_1,s_2)$ and $\psi(s_1,s_2)$
denote the deviations from the reference state density and composition, 
respectively. 
Using the fields introduced above, we propose the following Hamiltonian 
to govern the equilibrium shape of such an inhomogeneous membrane: 
\begin{eqnarray}
H&=& \int_\Omega dA \Big[\frac{1}{2}\kappa \left(\mathcal{C}-\mathcal{C}_0\right)^2+\sigma_0+\sigma_1  r\psi +\alpha_1\psi \left(\mathcal{C}-\mathcal{C}_{eq}\right) \nonumber\\
&+&\frac{\alpha_2}{2}\psi^2+\beta_1 r\left(\mathcal{C}-\mathcal{C}_{eq}\right)+\frac{\beta_2}{2}r^2\Big].
\label{HLG}
\end{eqnarray}
This functional includes the standard Helfrich 
energy density 
in the first two terms, with 
$\mathcal{C} (s_1,s_2)$=Tr($K_a^b$), being 
the mean local curvature of the surface, 
$\mathcal{C}_0$ and $\mathcal{C}_{eq}(s_1,s_2)$ are respectively the spontaneous and the equilibrium curvature of the reference state ($\psi$=0, $r$=0),
$\kappa$ the surface bending rigidity, and $\sigma_0$ the membrane surface tension. The shape of the reference state is obtained by minimizing the classical Helfrich Hamiltonian. %
Note that the Gaussian bending rigidity has been 
neglected 
here for simplicity. 
 
The subsequent terms were obtained %
by following a Landau-Ginzburg approach 
to take into account the effect of a variation in the membrane composition on its shape.  They include 
a self-energy contribution,
$\int(\alpha_2\psi^2/2+\beta_2 r^2/2+\sigma_1 r \psi) dA$,  depending only on the chemical state of the membrane and of phenomenological 
parameters $\alpha_2$, $\beta_2$, and $\sigma_1$; and terms 
that couple the geometry, {\it via} the curvature, 
 to the density, $\beta_1 r\left(\mathcal{C}-\mathcal{C}_{eq}\right)$, or 
to the compositional inhomogeneity, $\alpha_1\psi \left(\mathcal{C}-\mathcal{C}_{eq}\right)$. These terms can originate when  one considers lipids $L_1$ and $L_2$ of different shapes (see Fig. \ref{fig:1} for an illustration of such shapes).
The strengths of these couplings 
to the membrane shape are captured by the parameters $\beta_1$ and $\alpha_1$. %
Note that $\alpha_2, \beta_2>0$ for the energy to be well-defined, whereas 
$\alpha_1$, $\sigma_1$, and $\beta_1$ can be either positive or negative.

\subsection{Stress tensor, force balance, and the shape equation}

The stationary shape of a membrane described by a Helfrich-like Hamiltonian 
can be obtained from the shape equation, \ie the
force balance along the surface projected on the normal vector ${\bf n}$ (see Fig. (\ref{fig:1})). Following Guven and coworkers~\cite{CG02,Capovilla02, Guv04a}, we start with the derivation of the membrane stress tensor by introducing 
the constrained Hamiltonian 
\begin{widetext}
	\begin{eqnarray}
	\label{Hc}
	H_c&=&H+\int_{\Omega}\lambda^{ab}(g_{ab}-\mathbf{e}_a\cdot\mathbf{e}_b)dA+\int_{\Omega}\Lambda^{ab}(K_{ab}-\mathbf{e}_a\cdot\partial_b\mathbf{n})dA+\int_{\Omega}\mathbf{f}^a\cdot(\mathbf{e}_a-\partial_a\mathbf{X})dA \nonumber\\
	&+&\int_{\Omega}\lambda^a_{\bot}(\mathbf{e}_a\cdot\mathbf{n})dA+\int_{\Omega}\lambda_n(\mathbf{n}^2-1)dA.
	\label{eq:fullhamiltonianwithlagrangemultipliers}
	\end{eqnarray}
\end{widetext}
Here, the Lagrange multipliers $\lambda^{ab}$, $\Lambda^{ab}$, $\mathbf{f}^a$, $\lambda^a_{\bot}$, and 
$\lambda_n$ have been 
introduced to enforce the local definitions of the metric, the curvature tensor, the intrinsic basis, 
and the normal vector, respecively. % 

Previously, the 
Lagrange multiplier $\mathbf{f}^a$ has been identified as the stress tensor of the system~\cite{CG02,Capovilla02,Guv04a} . 
Its expression can be obtained via the variational minimization of $H_c$ with respect to the now  12 independent functions $\bf X$, $\mathbf{e}_a$,  $\mathbf{n}$, $K_{ab}$, $g_{ab}$, and  $r$. 
Note that we do not minimize with respect to the composition field $\psi$, 
which we consider to be an input of our model. We eventually obtain the following set of equations:
\begin{eqnarray}
	\label{eqLM1}
	{\bf X}:   \quad 0 &=&  \nabla_a\mathbf{f}^a  \\
	\label{eqLM2}
	\mathbf{e}_a: \quad 0 &=& - \mathbf{f}^a   +   ( \Lambda^{ac}K^b_c+2\lambda^{ab})\mathbf{e}_b    -\lambda^a_{\bot}\mathbf{n}\\
	\mathbf{n}: \quad 0 &=& ( \nabla_b\Lambda^{ab}+\lambda^a_{\bot})\mathbf{e}_a          + (2\lambda_n-\Lambda^{ab}K_{ab})\mathbf{n}\quad \\
	K_{ab}: \quad 0 &=&  \Lambda_{ab}  + \mathcal{H}^{ab}\\
	\label{eqLM5}
	g_{ab}: \quad 0 &=& \lambda^{ab}  -\frac{1}{2}T^{ab}  \\
	\label{eqLM6}
	r: \quad 0 &=& \beta_2 r+\frac{\beta_1}{ R^2}(\mathcal{C}-\mathcal{C}_{eq})+\sigma_1\psi.
	%\label{Eqr}
	%\label{eqLM7}
\end{eqnarray}

Here, $\mathcal{H}^{ab}$ and $T^{ab}$ are 
the functional derivatives of the Hamiltonian density ${\mathcal H}$ 
(defined according to $H=\int dA\ \mathcal{H}(s_1,s_2)$) 
with respect to the two fundamental forms $K_{ab}$ and $g_{ab}$: %
\begin{eqnarray}
\mathcal{H}^{ab}&=&\frac{\delta\mathcal{H}}{\delta K_{ab}}\\
T^{ab}&=&-\frac{2}{\sqrt{g}}\frac{\delta \sqrt{g}\mathcal{H}}{\delta g_{ab}}.
\end{eqnarray}
and $\nabla_a$ is the covariant derivative.
When $\mathcal{H}$ depends explicitly only on the local mean curvature $\mathcal{C}$, these derivatives can be expressed as
\begin{align}
\label{metstress}
T^{ab}&=-\mathcal{H} g^{ab} +2 \frac{\partial \mathcal{H}}{\partial \mathcal{C}} K^{ab} , \\
\label{curvstress}
\mathcal{H}^{ab}&=\frac{\partial \mathcal{H}}{\partial \mathcal{C}} g^{ab} .
\end{align}

%% new paragraph
From Eqs.~(\ref{eqLM1}-\ref{eqLM6}), one can obtain the % 
membrane stress tensor $\vb{f}^a$
\begin{equation}
	\label{stresstensor}
 \vb{f}^a=\qty(- \mathcal{H} g^{ab} - K^{ab} \frac{\partial \mathcal{H}}{\partial \mathcal{C}} )\vb{e}_b - \nabla_b \qty( \frac{\partial \mathcal{H}}{\partial \mathcal{C}} g^{ab})\vb{n}, 
\end{equation}
which expressed in the intrinsic basis is a $3\times2$ matrix. %

With this definition, in the absence of external forces, the force balance equations 
read,
\begin{align}
\label{ftan}
\nabla_a f^{a}_{\perp}- K_{ab} f^{ab} =&0,  \\
\label{fnorm}
\nabla_a f^{ab} + K^b_a f^{a}_{\perp} =&0,
\end{align}
where $f^{ab}=\qty(T^{ab}-\mathcal{H}^{ac} K_c^b )$, $f^{a}_{\perp}=  - \qty(\nabla_b \mathcal{H}^{ab})$, the tangential and normal components of the stress tensor such that $\vb{f}^a=f^{ab}\vb{e}_b + f_{\perp}^{a}\vb{n}$ (see Eq. (\ref{stresstensor})). 
Eq. (\ref{ftan}) corresponds to the force balance along the tangential directions.
The force balance along the normal direction, given in Eq.~(\ref{fnorm}), %
is generally referred to as the shape equation.

\subsection{Green's function for a tubular membrane and phase diagram of deformed shapes}

We now apply this framework to a cylindrical geometry and consider an infinite cylindrical membrane of radius $R$.  
We consider the reference state to be  
a cylinder with 
$\psi_0=0$, $r_0=0$.  For vanishing inhomogeneities, the system is described by the standard Helfrich model. Note that  if  $\psi=0$ and $\mathcal{C}=\mathcal{C}_{eq}$,  then $r=0$ follows from equation Eq.~\ref{eqLM6}. Using Eqs.~(\ref{metstress}-\ref{stresstensor}), %
we derive the tangential stress tensor for the reference state. We express
it in the intrinsic coordinates: $\theta=s_1$, the revolution angle, and $s=s_2$, the arclength, associated with the intricinc basis (${\bf e}_\theta, {\bf e}_s $, ${\bf n} $). Note that this coordinate system coincides with the cylindrical basis for the undeformed state.  It reads 
\begin{eqnarray}
\label{f0}
f_{0}^{ab}&=&\left(\begin{array}{cc} \frac{ \kappa(1-X^2)-2\sigma_0R^2}{2R^4}  & 0\\
0 &    -\frac{ \kappa (1-X)^2}{2R^2}-\sigma_0 \end{array}\right),
\end{eqnarray}
where we  have introduced the dimensionless parameter 
\begin{equation} %
X=R \mathcal{C}_0 .%
\end{equation} %
The normal part of the stress tensor, $(f^{\theta}_{\perp0}$, $f^{s}_{\perp  0}$) vanishes. 

The force balance in the direction ${\bf e}_{\theta}$,
Eq.~(\ref{ftan}), vanishes for symmetry reasons. The remaining force balance, along the normal direction, for the undeformed cylinder, 
Eq.~(\ref{fnorm}), reads %
\begin{equation}\label{eq:F_p_0}
\frac{1}{2}\frac{\kappa}{R^3}(1-X^2)-\frac{\sigma_0}{R}=0. %
\end{equation}

  Assuming positive values for the tension $\sigma_0$, the domain of stable solutions for Eq. (\ref{eq:F_p_0}) is 
$X \in  ]-1,1[$
and one 
eventually 
obtains the following solution for the equilibrium radius $R$ in the reference state,
\begin{equation}
R =\frac{1}{\mathcal{C}_{eq}} = \frac{1}{\sqrt{\mathcal{C}_0^2 + 2 \sigma_0/\kappa}},
\end{equation}
where we have used the knowledge that the curvature of a cylinder (here $\mathcal{C}_{eq}$) is $1/R$. Note the curvature at the equilibrium equals the spontaneous curvature only for vanishing tension.
We use this relation to express the bending rigidity $\kappa$ in terms of $R$, $\sigma_0$, and $X$ in the following.

We next derive the shape of a deformed cylinder in response to a 
ring-like perturbation of the composition obeying the rotational symmetry of the cylinder, %
$\psi(\theta,s)=\psi(s)$, with $\psi(s)=\delta\psi \delta(s)$, where $\delta\psi$ is a magnitude and $\delta(s)$ a Dirac delta for the surface coordinate $s$, see Fig.~\ref{fig:1}b. %
To describe the deformations induced by the perturbation, we 
introduce the deformation field $\vb{u}(s)=u_n(s) \vb{n} + u_s(s) \vb{e}_s$ % (from figure legend)
with radial and tangential components  $u_n(s)$ and  $u_s(s)$, respectively % 
(see Fig.~\ref{fig:1}b), where $s$ is now the acrlength associated with the deformed shape. %
The differential geometry elements associated with the deformed tube, 
 \ie the intrinsic basis 
${\bf e}_\theta$, ${\bf e}_s$, $n$, as well as the metric and  mean curvature of the surface, are given in the Appendix A. We derive them  up to first order in the response fields,  \ie the deformation fields ($u_n, u_s$) and the density inhomogeneity 
$r$, which are assumed to be small - \ie $u_n/R$, $u_s/R$ and $r \ll 1$.

The surface stress tensor in the deformed state writes $f=f_{0}+f_{1}$ and $f_{1}^{ab}$ %
can be decomposed into a mechanical part that depends on $(u_n, u_s)$
and a chemical part that is a function of ($r$, $\psi$) 
such that $f_{1}^{ab}=f_{1,M}^{ab}+f_{1,C}^{ab}$, where 
\begin{widetext}

\begin{eqnarray}
	\label{eq:fab_1M}
	f_{1, M}^{ab}&=&\left(\begin{array}{cc}  -\frac{2 \sigma_0}{R^3(1-X^2)}(u_n+XR^2u_n'')  & 0\\
		0 &    \frac{2 \sigma_0}{R(1+X)}u_n+\frac{4\sigma_0}{1+X}u_s'  \end{array}\right), \\
	f_{1, C}^{ab}&=&\left(\begin{array}{cc}  \frac{1}{R^3}\qty(\beta_1 r + \sigma_1 \psi)& 0\\
		0 &  0   \end{array}\right) . %
\end{eqnarray}
\end{widetext}
The same decomposition can be applied to the normal part of the 
stress tensor, $f^{a}_{\perp} = f^a_{\perp 1,M} + f^a_{ \perp 1,C}$,  with 
\begin{eqnarray}
f_{\perp 1,M}^{a}&=&\left(0,\quad \frac{2 \sigma_0}{1-X^2}\qty(u_n'+R^2u_n''')  \right) \\
\label{eq:fan_1C}
f_{\perp 1,C}^{a}&=&\left(0,\quad -\beta_1 r'-\alpha_1 \psi'\right).
\end{eqnarray}
As can be seen from Eqs.~(\ref{eq:fab_1M}-\ref{eq:fan_1C}), %
the coupling of the membrane curvature and the internal degrees of freedom $r$ and $\psi$ 
leads to extra lateral tensions and extra bending forces proportional to the coupling coefficients. 
Using the expression of the stress tensor and 
the force balance along ${\bf e}_s$ and ${\bf n}$, using $r(s) = - \frac{\beta_1}{\beta_2 R^2}\qty(\mathcal{C}(s)-\frac{1}{R})-\frac{\sigma_1}{\beta_2}\psi(s)$ according to Eq.~(\ref{eqLM6}) in the force balance along ${\bf n}$, one gets the following shape equation for the system:
\begin{widetext}
	\begin{eqnarray}
\frac{\beta_1^2}{\beta_2 R^2}(u_n+2R^2u_n''+R^4u_n'''')-\frac{2 \sigma_0}{1-X^2}(u_n+2XR^2u_n''+R^4u_n'''')&=&\left(\alpha_1-\sigma_1\frac{\beta_1}{\beta_2}\right)(\psi+R^2\psi'') . 
			\label{Eqfn}
	\end{eqnarray}
\end{widetext}
The left-hand part is the differential equation controlling the normal deformation field $u_n$, the right-hand part corresponds to the source term in $\psi$.
One sees here the predominant role played by the coupling between density and curvature 
proportional to $\beta_1$.  The prefactor of the  higher-order derivative term in $u_n''''$ is equal to 
$(\beta_1^2/\beta_2-\kappa)R^2$. % 

For a non-vanishing coupling ($\beta_1\neq 0$), %
one can define an effective bending constant based on 
the prefactor of  $u_n''''$ in the shape equation (\ref{Eqfn}).
Let us now define the Green's function of the system, $G_{\perp}(s)$, 
according to 
$u_n(s)=\int ds' G_{\perp}(s-s')\psi(s')$. 
Performing a Fourier transform of Eq.~(\ref{Eqfn}), %%~
one gets 
\begin{widetext}
	\begin{equation}\label{eq:Green_q_raw}
	\frac{\tilde{G}_{\perp}(q)}{R} = \frac{\delta \psi}{2\sqrt{2\pi}}  \frac{ Z (1-X^2) (R^2 q^2 - 1)}{ [Y(1-X^2) - 1] R^4 q^4   -  2[Y(1-X^2) - X] R^2 q^2 + Y(1-X^2) - 1 } , %%, and outer () -> []
	\end{equation}
\end{widetext}
where we have furthermore introduced the two additional dimensionaless parameters 
\begin{equation}
Y=\frac{\beta_1^2}{2\sigma_0\beta_2 R^2}\quad \text{and}\quad   Z=\frac{1}{R \sigma_0}\qty(\alpha_1-\sigma_1 \frac{\beta_1}{\beta_2}). %, %% .
\end{equation}
Note that we used the  convention of the Fourier transform 
 $\tilde{u}(q)=1/\sqrt{2\pi}\int ds e^{iqs} u(s)$. 
 
 \begin{figure}
	\includegraphics[scale=1.45]{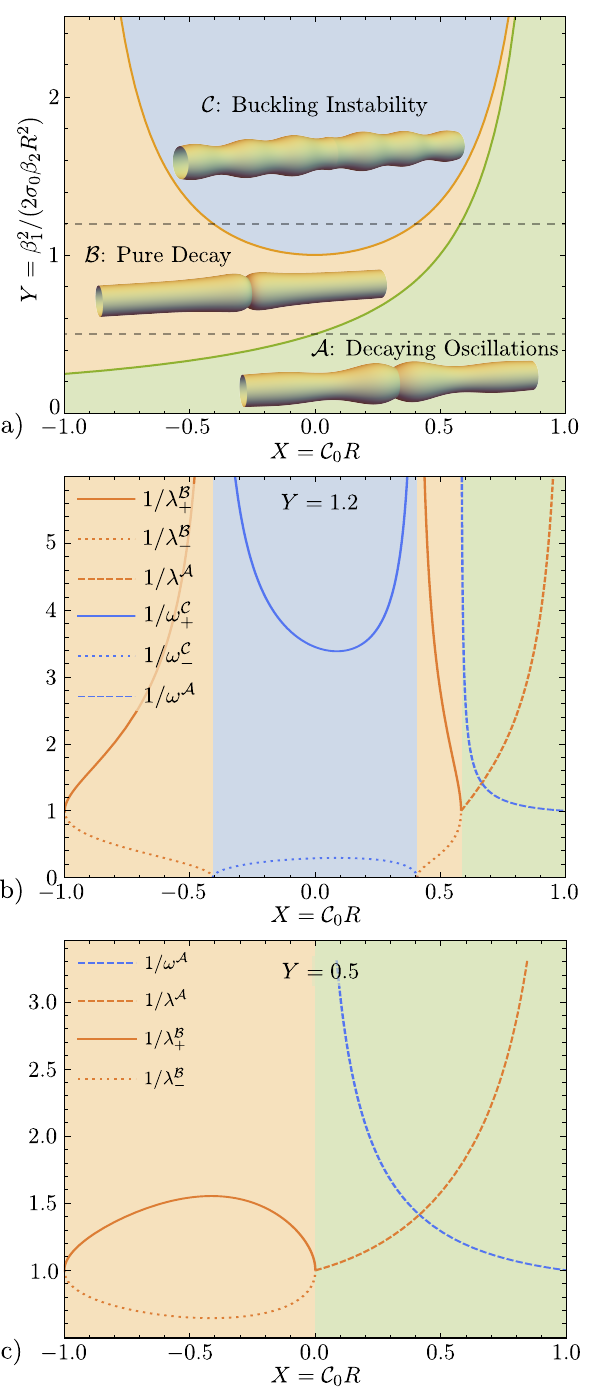} % AR : 0.6
	\caption{a) Phase diagram of the deformed cylinder in the ($X$, $Y$) plane.  Region $\mathcal{A}$ (green): decaying oscillation deformation; region $\mathcal{B}$ (orange), decayig deformation; region $\mathcal{C}$ (blue): buckling instability.  Decaying $(R/\lambda)$ and oscillation $(R/\omega)$ lengths as a function of $X$ for (b) $Y=0.5$ and (c) $Y=1.2$. }
	\label{fig:2}
\end{figure}

The Green's function is a function  of 
the three dimensionless parameters $X$, $Y$, 
and $Z$ that obey $-1<X<1$,  $Y\geq0$ and $Z$ unbounded.  
While $X$ 
(defined 
in Eq.~(\ref{f0})) depends only on the geometric parameters,
 $Y$ and $Z$ are non-trivial functions of the parameters introduced in the Landau-Ginzburg expansion of the Hamiltonian given in Eq.~(\ref{HLG}). %%~
$Z$ determines the amplitude and sign of the response. The four poles, ($\pm q_+, \pm q_-$),  of $G_{\perp}(s)$ are functions, of $X$ and $Y$
(see expression in the Appendix B).

The Green's function allows us to study the phase space of the model, 
i.e., the shape of the mechanical response $u_n$ as a function of $X$ and $Y$. The shape phase diagram is shown in  Fig.~\ref{fig:2},  
and presents three distinct regions defined as 
follows. %
In region $\mathcal{A}$ (shown in green in Fig.~\ref{fig:2}a) %
the roots are complex with non-vanishing %
real and imaginary parts, %
\begin{equation}
\mathcal{A}: \quad 0< Y < \frac{1}{2}\frac{1}{1-X}; \quad R q_{\pm}=\omega^{\mathcal{A}} \mp i \lambda^{\mathcal{A}}
\end{equation}
In this region, deformations oscillate with spatial period (oscillation length) $2\pi R/\omega^\mathcal{A}$ and decay over a characteristic %
decay length $R/\lambda^{\mathcal{A}}  > 0$. %

The Green's function in the real space reads 
\begin{eqnarray}
G_{\perp}^{\mathcal A} \qty(s) &=& { \Upsilon} \frac{e^{-\lambda^{\mathcal{A}}\frac{\qty|s|}{R}}}{2} \Bigg[\frac{1}{\lambda^{\mathcal{A}}} \qty(1-\frac{1}{\lambda^{\mathcal{A}\,2}+\omega^{\mathcal{A}\,2}})\cos\qty(\omega^{\mathcal{A}} \frac{\qty|s|}{R} )\nonumber \\
&-& \frac{1}{\omega^{\mathcal{A}}} \qty(1+\frac{1}{\lambda^{\mathcal{A}\, 2} + \omega^{\mathcal{A}\, 2}}) \sin\qty(\omega^{\mathcal{A}} \frac{\qty |s|}{R})    \Bigg],
\end{eqnarray}
where $\Upsilon=\frac{\delta \psi}{4}  \frac{1-X^2}{Y\qty(1-X^2)-1} Z $. The tube plotted in region $\mathcal{A}$ (Fig.~\ref{fig:2}a) %
gives an illustration of the shapes obtained in this zone.

For region $\mathcal{B}$  (shown in orange in Fig.~\ref{fig:2}a) the poles are purely imaginary, %,
\begin{equation}
	\mathcal{B}: \quad \frac{1}{2}\frac{1}{1-X} < Y < \frac{1}{1-X^2}; \quad R q_{\pm} = i \lambda_{\pm}^{\mathcal{B}},
\end{equation}
 with $\lambda_{+}^{\mathcal{B}}>\lambda_{-}^{\mathcal{B}}>0$, and deformations are associated with two decay lengths $R/\lambda^{\mathcal{B}}_{\pm}$, according to %
\begin{equation}
G_{\perp}^{\mathcal B}\qty(s)=  \frac{\Upsilon}{\lambda_+^{\mathcal{B}\,2}-\lambda_-^{\mathcal{B}\,2}}  \Bigg( \frac{\lambda_+^{\mathcal{B}\,2} + 1}{\lambda_+^{\mathcal{B}}} e^{-\lambda_+^{\mathcal{B}} \frac{\qty|s|}{R}} -  \frac{\lambda_-^{\mathcal{B}\,2} + 1}{\lambda_-^{\mathcal{B}}} e^{-\lambda_-^{\mathcal{B}} \frac{\qty|s|}{R}}\Bigg).
\end{equation}

%% new paragraph
Finally, the region $\mathcal {C}$ (blue in Fig.~\ref{fig:2}a) %
is  associated with  four real poles,
\begin{equation}
\mathcal{C}:  \quad \frac{1}{1-X^2} < Y, \quad R q_{\pm}=\omega_{\pm}^{\mathcal C}>0 . %.
\end{equation}
This region corresponds to a buckling instability  
zone. Competition between tangential and normal forces in tubular membranes  can produce peristaltic shapes which wavelength depends on the system\cite{bar-ziv94}. Here, a local perturbation induces a non decaying response 
that is given by %
\begin{eqnarray}
G_{\perp}^{\mathcal C}(s)&=& \frac{\Upsilon}{\omega_+^{{\mathcal C}\,2}-\omega_-^{{\mathcal C}\,2}} \Bigg[ \frac{1-\omega_+^{\mathcal{C}\,2} 
}{\omega_+^{\mathcal{C}}} \sin\qty(\omega_+^{\mathcal{C}}\frac{\qty|s|}{R}) \nonumber\\ &-& \frac{ 1 - \omega_-^{\mathcal{C}\,2} }{\omega_-^{\mathcal{C}}} \sin\qty(\omega_-^{\mathcal{C}}\frac{\qty|s|}{R}) \Bigg],
\end{eqnarray}
which involves two oscillating lengths 
$2\pi R/\omega_{\pm}^{\mathcal{C}}$.   %

The characteristic lengths - the inverses of real and imaginary parts of the poles given above - rescaled by the cylinder radius $R$, are plotted in 
Fig.~\ref{fig:2}b,c as functions of $X$ for two fixed values of $Y$. 
When $Y=0.5$ (Fig.~\ref{fig:2}c), one sees that for $X<0$, corresponding to region $\mathcal{B}$, %
the two decay lengths $1/\lambda_\pm^{\mathcal{B}}$ %
(shown in  orange solid and dotted lines) remain finite with $1/\lambda_{+}^{\mathcal{B}} > 1$ ( orange solid line) and $1/\lambda_{-}^{\mathcal{B}}<1$ for $-1<X<0$, while both take the value of $1$ at $X=-1$ (the edge of the valid domain for X) and $X=0$.
For $X>0$, \ie in region $\mathcal{A}$, %
the oscillation length $1/\omega^{\mathcal{A}}$ 
(blue dashed line) diverges at $X=0$ but 
decreases monotonically with $X$ %
to reach $1$ for $X=1$. %
On the contrary, the decay length $1/\lambda^{\mathcal{A}}$  
(red dashed line) %
monotonically increases from $1$ for $X=0$ to diverge in the  limit $X\rightarrow 1$, %,
indicative of a buckling instability 
at $X=1$. 
Fig.~\ref{fig:2}b shows the characteristic lengths as a function of $X$ for $Y=1.2$. 
The behavior of the characteristic lengths in 
regions $\mathcal{A}$  and $\mathcal{B}$ is 
similar to the one 
previously described, but a  
domain pertaining to region 
$\mathcal{C}$ is intercalated in 
region $\mathcal{B}$ 
(at $-0.48<X<0.48$), domain in which the oscillation %
length $1/\omega_+^{\mathcal{C}}$ 
(blue solid line) %
reaches a minimum between its two positive diverging limits  
on the
borders between regions $\mathcal{B}$ and $\mathcal{C}$. %
Finally, $1/\omega_-^{\mathcal{C}}$ 
(blue dotted line) vanishes on the 
%$\mathcal{B}/\mathcal{C}$ boarder 
borders between regions $\mathcal{B}$ and $\mathcal{C}$ %
and reaches a maximum in between. % space
The analysis of the Green's functions on the transition lines is presented 
for completeness %
in the Appendix C.

To conclude, in this section %
we have derived the shape equation for an inhomogenous membrane and 
%calculated 
determined %
the phase diagram of 
%this system
the model % 
in 
%an 
a %
cylindrical geometry. We have shown that the coupling of geometry and internal degrees of freedom 
can %
generate a rich variety of responses. In the next section, we apply this model to the mitochondrial 
inner %
membranes, %.
or cristae.

\section{Application to the cristae membrane}

In this section, we use our model to describe the deformation of the mitochondrial cristae induced by a surface 
pH gradient. % 

Mitochondria are composed of two membranes, the 
so-called outer and the inner membrane. While the outer membrane provides the outer envelope of the mitochondrion (as its name suggests), the inner membrane (IM) - enclosing the matrix - contains the protein complexes necessary for energy production. More specifically, the IM %
forms tubular and  ``pancake-like'' %
invaginations named cristae that are the place of ATP synthesis from ADP. This endothermic reaction is catalyzed %
by ATP synthase, %
a transmembrane protein located in the 
%highly curved zones 
zones of high curvature %
of the cristae. 
%ATPsynthase 
ATP synthase %
uses a gradient of 
the %
proton electrochemical potential between the two sides of the IM
%protomotrice 
driving %
force. 
%It 
 Protons in the cristae are supplied by transmembrane proteins of the 
electron transport %
chain 
(also referred to as respiratory chain), %
which inject protons from the matrix 
in the cristae. As these %
proteins are located in flat regions %
of the cristae, ATP synthase %
and respiratory chain proteins are spatially separated and the protons  diffuse from one site to the other. Notably, the %
protons are thought to diffuse along the cristae membrane 
%-an
and not in the bulk%
%of the invagination -   
~\cite{heberle1994,heberle2000,gennis2016}. % ~, space

The mechanisms 
that couple respiratory chain complexes and ATP synthase proteins to ensure efficient ATP synthesis are a very active field of research. 
%In this part, 
Here, %
we focus on the coupling between 
%the 
a %
proton gradient and the 
%membrane 
shape of the cristae 
membrane %
which has been observed to 
%go 
change % 
from a regular cylinder to a bumpy irregular tubule
%, 
when submitted to proton gradients of increasing intensity~\cite{mannella2006structure}. %~

\subsection{Cristae membrane composition and the importance of cardiolipins}

The mitochondrial IM is partly composed of cardiolipin (CL), an anion-acid lipid, found in the majority of membrane organelles 
that enclose oxidative phosphorylation processes. %
The value of CL's 
second pKa, associated with the proton exchange,
\begin{equation}
	\label{eq:chemical}
{\rm CL}^{2-}+H^+\rightleftharpoons{\rm CLH}^- \quad {\rm pKa}=8
\end{equation}
suggests that at physiological conditions - pH=7 -, the CL carries a unique charge~\cite{haines2002}. %~

It has been speculated that this % 
lipid could 
thus %
play 
%a 
the %
role of a proton trap creating a surface proton reservoir independently of the %bulk-cristae 
bulk %
pH 
in the cristae%
~\cite{heberle2000,haines2002}. % ~ and space
Moreover, %, 
{\it in vitro} experiments suggest that the change of 
the %
protonation state of CL could affect the membrane mechanical properties~\cite{khalifat2008membrane}. %~

In order to take into account the potential coupling of the CL protonation state (related to the local proton concentration) to the membrane shape, we 
thus consider 
the density field $\rho$ and the composition field $\phi$ introduced in the model (see Eq.~\eqref{HLG}) %
%as 
to be given by %
the 
%density 
densities %
of 
the %
two cardiolipin forms HCL$^-$ and CL$^{2-}$, 
such that % 
$\rho(s)=\rho_{\rm CL^{2-}}+\rho_{\rm HCL^{-}}$ and 
%$\phi(s)=\frac{\rho_{\rm CL^{2-}}}{\rho}$
$\phi(s)={\rho_{\rm CL^{2-}}}/{\rho}$. %

In the following, we model the cristae as a 
membrane 
tube %
of length $L$, 
that is %
closed by a spherical cap which we will not further consider here. %
The reference state corresponds to a homogeneous CL distribution in the absence of catalytic activity of the electron transport chain and ATP synthase, and accordingly $\psi=0$. %
We assume that the 
(cylindrical) %
reference state is %
in mechanical equilibrium. 
Its internal tension is counterbalanced by an external tension ${\bf f}_{\rm ext}$ applied at $s=0,L$.
Following Eq.~{\ref{f0}}, this 
external force 
is given by 
\begin{equation}
 {\bf f}_{\rm ext}(0)=-{\bf f}_{\rm ext}(L)=-\frac{2 \sigma_0}{1+X}{\bf e}_s. 
\end{equation} 

 \begin{figure*}
	\includegraphics[scale=1.0]{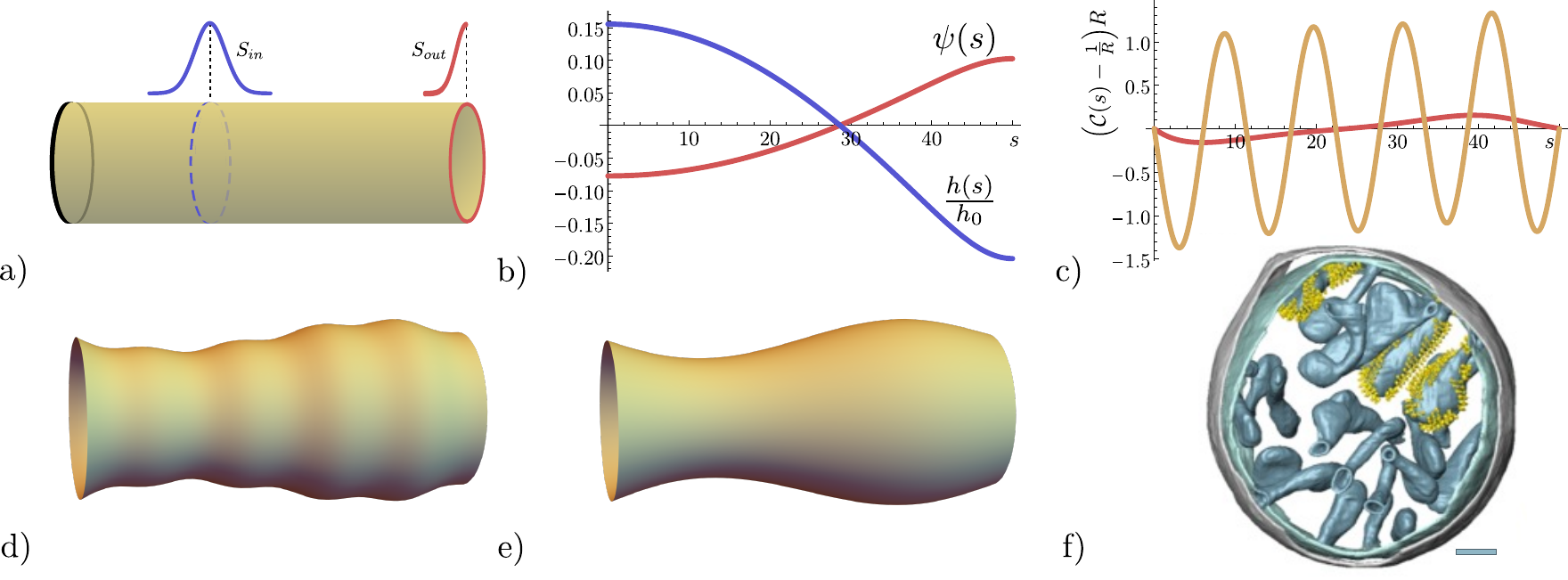} % AR : 0.6
	\caption{a) Finite cylinder of length $L$ on which is included a surface source $S_{\rm in}$ and sink $S_{\rm out}$ of protons. b) Proton profile $h(s)/h_0$  and consequent lipid composition $\psi(s)$. Solutions of Eqs. (\ref{eq::h_dif}, \ref{compo}) for parameters given in Tab. (\ref{tab:values}). c) Contribution to the density inhomogeneity proportional to the curvature $(\mathcal{C}-\mathcal{C}_{eq})$ (plot red corresponds to shape e) plot orange to shape d) ).  d) Example of a solution of Eq. (\ref{Eqfn}) for $h_0=30$ protons.nm$^{-1}$, $\phi_0=0.5$, $X=0.21$, $Y=1.1$ and  $Z=17$. e) Example of a solution for the same parameters as d) except for $X=-0.25, Y=0.5,  Z=40$. f) Reproduction of a graphical scheme of a Cryo-ET of isolated Polytomella sp. mitochondria, scale bar $100$nm.\cite{blum2019}}
	\label{fig:3}
\end{figure*}

In order to determine the compositional profile $\psi(s)$ of the CLs in the catalytically active state, we first consider the surface concentration gradient of protons along the cristae membrane and then infer the local shift in CL$^{2-}$ and  HCL$^-$ concentrations based on the protonation rate constants. %
The surface concentration profile of protons can be written as 
$[\text{H}^+]=h_0+h(s)$. %
The homogeneous proton concentration $h_0$, associated with the reference state,  corresponds to the density of protons trapped by 
the CL  in the undeformed cylinder,
for which both the respiratory chain proteins and the 
ATP synthase do not present catalytic activities. Note that this state is also referred to as 
state IV of the organelle~{\cite{mannella2006structure}}. % cite 
We denote by %
$h(s)$ 
%is 
the perturbation of the surface proton field due to the catalytic activities of the proteins that inject and consume H$^+$. 
To 
%derive 
determine %
$h(s)$, we model the system as follows (see sketch
in %
% Fig. 3 a.): 
Fig.~\ref{fig:3}a). %
%at 
At one end of the cylinder ($s=0$), %, 
%one finds 
we assume %
a reflecting barrier for the proton flux, $dh/ds|_{s=0}$, % 
modeling the role of the cristae junction proteins~{\cite{vanderlaan2016}}. % cite
At the other end of the cylinder
%, $s=L$ 
($s=L$), %
a ring-shaped 
%of 
proton sink
%, associated 
with a spatial extension of $\Delta_{\rm out}$ %, 
models the proton consumption of %catalytic activity of the % 
ATP synthase proteins.
At $s=L_s$, a ring-shape proton source, 
%associated 
with a spatial extension of $\Delta_{\rm in}$, models the 
proton insertion of the %catalytic activity of the %
respiratory chain. The profile $h(s)$ 
then %
satisfies the following stationary diffusion equation, %,
\begin{equation}\label{eq::h_dif}
D \frac{\partial^2 h}{\partial s^2} \qty(s,t) + S_{\rm in}\qty(s) - S_{\rm out}\qty (s)=0,
\end{equation}
where 
$D$ is the proton diffusion constant and %
\begin{eqnarray}
S_{\rm in} \qty(s) &=& \frac{k_{\rm in}}{\sqrt{2\pi \Delta_{\rm in}^2}} \exp \qty(-\frac{1}{2\Delta_{\rm in}^2} \qty(s-L_s)^2) \\ S_{\rm out} \qty(s) &=& \frac{k_{\rm out}  (h_0+h(s))}{\sqrt{2\pi \Delta_{\rm out}^2}} \exp \qty(-\frac{1}{2\Delta_{\rm out}^2} \qty(s-L)^2)\quad
\end{eqnarray}
are the expressions for the source and sink  
injection and consumption 
rates, respectively. %. 
The numerical values of the parameters introduced here are 
%given 
specified %
in Table~\ref{tab:values}. % no ()

%%%%%%%%%%%

%We solve 
The ratio $h(s)/h_0$ - with $h(s)$ solution of %
Eq.~(\ref{eq::h_dif}) - %
%and plot $h(s)/h_0$ in Fig. 3b. 
is shown in Fig.~\ref{fig:3}b (blue line). %
The proton concentration profile indicates an excess of protons
in the zone of the respiratory chain (modeled by the proton source) %
%in the respiratory chain zone 
and a lack of protons in the 
%sink 
zone 
of ATP synthase enrichment (modeled by the proton sink), %
in agreement with {\it in vivo} pH measurements%
~\cite{rieger2014lateral}. %~

%% new paragraph
The inhomogeneity in the proton profile will induce an inhomogeneity in the lipid composition 
%.
due to a shift in the chemical equilibrium between the two forms of CL. %
%The 
In the %
reference state, %,
%is defined as 
the lipid composition is 
%$\phi_0=\frac{\rho^0_{\rm CL2-}}{\rho^0}.$ 
$\phi_0=\rho_{{\rm CL^{2-}},0}/\rho_0$.
Using the conservation of matter, the variation in the composition can be expressed 
to %
the first order in $h$ as %,
\begin{equation}
\label{compo}
	\psi(s)=-(1-\phi_0)\frac{h(s)}{h_0}. %.
\end{equation}
See Appendix D for details. The profile $\psi(s)$ for an initial $\phi_0=0.5$ is plotted in 
Fig.~\ref{fig:3}b (red line). As expected, one can observe % 
an enrichment in protonated  
CLs (corresponding to $\psi<0$) %
close to the 
proton %
source 
%$\psi<0$ 
and 
%of 
in %
deprotonated 
%ones 
CLs %
close to the 
proton %
sink. 

\subsection{Mitochondrial cristae shape in the presence of catalytic activity}

Based on our simple model of a CL composition gradient in the cristae membrane, we %
now solve the shape equation 
Eq.~(\ref{Eqfn}) 
given the inhomogeneous field 
$\psi(s)$ %, 
and assuming that the mechanical boundary conditions remain unchanged, 
\ie we still consider 
a pinned protrusion with 
fixed (reference state) %
curvature in $s=0$ and $s=L$. 

The latter leads %
to the following conditions on the radial deformation 
field: % 
\begin{eqnarray}
	u_n(0)&=&0, \quad u_n(L)=0\nonumber\\
	u''_r(0)&=&0,  \quad u_n''(L)=0.
\end{eqnarray}
The solutions to the shape equation subject to these boundary conditions were obtained by numerical integration.

\begin{table*}
	
	\renewcommand{\arraystretch}{1.8}
	\centering
	
	\begin{tabular}{|c|c||c|c|}
		\cline{1-4}
		$L$ & $50\ \si{\nano \meter}$                               & $k_{\rm in}$ & $6\ 10^6$\ $\si{\second^{-1}}$ \cite{rieger2014lateral}  \\                    
		$R$ & $10\ \si{\nano \meter}$                               &  $k_{\rm out}$ & $3.24\ 10^5\ \si{proton^{-1} s^{-1}} \cite{rieger2014lateral} $   \\
		$L_s$ & $9.5\ \si{\nano\meter}$ & $D$ & $10^7\ \si{\nano \meter^2 \second^{-1}}$ \cite{gennis2016}\\
		$\Delta_{\rm in}$ &  $20\ \si{\nano\meter}$ & $\Delta_{\rm out}$ & $5\ \si{\nano\meter}$ \\ 
		$\kappa$ & $10^{-10}\ \si{\newton\ \nano\meter} $       &
		$\sigma_0$ & $10^{-16}\ \si{\newton\ \nano\meter^{-1}}\cite{AF2}$  \\		
		\hline	 
	\end{tabular}
	\caption{Values of the model parameters.}
	\label{tab:values}	
\end{table*}

In Fig.~\ref{fig:3}d,e, we show the membrane shapes 
obtained for two  sets of parameters, one belonging to 
region $\mathcal{B}$ 
of the shape phase diagram (Fig.~\ref{fig:3}e) %
and one to region $\mathcal{C}$ 
(Fig.~\ref{fig:3}d).
Both present a  bulge close to the proton sink and a narrow bottleneck around the proton source. 
For the parameters belonging to region $\mathcal{B}$ (Fig.~\ref{fig:3}e), the overall shape is smooth, whereas for parameters of region $\mathcal{C}$ (Fig.~\ref{fig:3}d), the shape presents a succession of oscillations modulating the 
major bulge and narrow, %
signature of the buckling 
instability
observed in this region. %
%These 
In general, the %
shapes are in qualitative agreement with recent 
%electronic 
electron %
microscopy observations. A Cryo-ET image of mitochondry is reproduced from~\cite{blum2019} in Fig.~\ref{fig:3}f and one can identify  the necks and bulges in cristae. Finally, we consider the density inhomogeneity $r(s)$ induced by the composition perturbation $\psi(s)$.  The expression given in 
Eq.~(\ref{eqLM6}) %~
shows that $r(s)$ is the sum of a term proportional to $\psi$ and a term involving the curvature. When the coupling between $r$ and $\psi$ dominates, quantified by the value of $\sigma_1$, the inhomogeneity in mass density follows $\psi$. When the coupling between $r$ and the curvature is dominant, quantified by the value of $\beta_1$ and consequently $Y$, the 
inhomogeneity %
in the density will be given by the variation of the curvature $(\mathcal{C}-\mathcal{C}_{eq})$. This contribution 
is plotted in Fig~\ref{fig:3}c 
for the two shapes we have considered 
above and shown in Fig.~\ref{fig:3}d,e. One retrieves the unique 
narrow %
and 
%bump 
bulge %
of the 
%form
shape % 
shown in 
Fig~\ref{fig:3}e (red line in Fig~\ref{fig:3}c) %
and a succession of dense and sparse zones
remindful of the shape shown %
in 
%the form Fig 3. d) (orange plot).
Fig~\ref{fig:3}d (orange line in Fig~\ref{fig:3}c). %

\section{Conclusion}

In this work, %,
we 
first %
presented a generalized Helfrich model for 
inhomogeneous membranes %
%that we couple to a diffusive flux driving inhomogeneity on the surface. 
with a coupling between membrane geometry and internal degrees of freedom related to membrane composition. %
We showed that the shape of such systems can easily be studied after derivation of the stress tensor obtained via a constrained minimization of the Hamiltonian. We show that, in 
%thee 
the %
case of cylindrical geometries, these systems present a rich phase diagram and could explain deformations observed both for {\it in vitro} and {\it in vivo } systems. 
This model could be applied to CL membranes that are known to deform under pH variation \cite{khalifat2008membrane}. Recently, controled microfluidic devices were developed to monitor the vesicles response to a variation of chemical environement (salt concentration or pH)~\cite{karimi2018,pramanik2022}. Such protocols could be used to validate and parametrize the model.

We 
%apply 
then applied %
this framework, %,
combined with a simple model of proton transport at the cristae surface, % 
to describe the 
shape deformations 
of mitochondrial cristae driven by the surface proton flux established between proton 
sources and sinks, %
and that arise as a consequence of inhomogeneities in the membrane composition downstream of the proton concentration gradient. % 
%We recover 
Our model reproduces %
the 
%alterning 
characteristic alternation between more constricted and wider regions 
%condensed and bumpy shapes characterizing
typically associated with %
%low and high 
spatially varying rates of 
%ATP-synthesis rates of 
ATP synthesis in %
the cristae~\cite{blum2019}. 

%% TBD: 
%% --------
%% - appendix
%% - figure legends 

\paragraph*{Acknowledgments}
H.B.~thanks A.-F.~Bitbol, J.~Heberle, S.~Bloch and R.~Netz for usefull discussions.
H.B.~acknowledges funding from Humboldt Research Fellowship Program for Experienced Researchers.

\section{Appendix}
\subsection{Elements of differential geometry for a deformed cylinder}
We use a standard parametrization for axisymmetric surfaces, $s$ is the arclength, $\theta$ the revolution angle.  $\vb{X}(\theta,s)$ gives the cylindric surface and the deformation components are $\delta \vb{X}=\vb{u}$. 
We assume no deformations on the $\theta$ direction. Thus it gives:

\begin{equation}
\vb{X}=\begin{pmatrix} R \cos{\theta} \\ R \sin{\theta} \\ s \end{pmatrix}\ ,\ \delta \vb{X}=\begin{pmatrix} u_n(s) \cos{\theta} \\ u_n(s) \sin{\theta} \\  u_s(s) \end{pmatrix}
\end{equation}
where  $\theta \in [0,2\pi[$ and $s \in [0,L]$, these are the curvilinear coordinates on the surface of the undeformed cylinder of radius $R$ (see Figure \ref{fig:1}).
The intrinsic basis is given by:
\begin{eqnarray}
\vb{e}_{1}&=&\vb{e}_{\theta}=\begin{pmatrix} -\qty(R+u_n) \sin{\theta} \\ \qty(R+u_n) \cos{\theta} \\ 0 \end{pmatrix},\nonumber \\
\vb{e}_{2}&=&\vb{e}_{s}=\begin{pmatrix} u_n' \cos{\theta} \\ u_n' \sin{\theta} \\ 1+u_s'\end{pmatrix}, \nonumber \\ \quad\vb{n}&=&\begin{pmatrix} \cos{\theta} \\  \sin{\theta} \\ -u_n' \end{pmatrix}.
\end{eqnarray}
where we use $s_1=\theta$ and $s_2=s$ such that the normal vector points to the outside of the tube and the prime means derivative with respect to $s$. 
The metric $g_{ab}$ and the curvature $K_{ab}$ are expressed to the first order in the deformation field 
\begin{equation}
g_{ab}=\begin{pmatrix}
R^2+2 R u_n  & 0 \\
0 & 1+2 u_s'
\end{pmatrix}, \quad 
K_{ab}=\begin{pmatrix}
R + u_n  & 0 \\
0 & -u_n''
\end{pmatrix},
\end{equation}
 where $a$ and $b$ are equal to $1$ or $2$. With this parameterization of our system the area element is given by:
\begin{equation}
dA=(R+u_n+R  u_s')d\theta ds.
\end{equation}
Moreover, the normal and scalar curvatures are given by:
\begin{equation}\label{eq:cyl_C_R}
\mathcal{C}=\frac{1}{R} - \qty(\frac{u_n}{R^2}+u_n''),\quad \mathcal{R}=- 2 \frac{u_n''}{R},
\end{equation}
and the Christoffel symbols are
\begin{equation}
	\Gamma^{\theta}_{ab}=\begin{pmatrix}
		0  &  u_n'/R \\
		u_n'/R   &  0
	\end{pmatrix}, \quad 
	\Gamma^{s}_{ab}=\begin{pmatrix}
		-R u_n'  & 0 \\
		0 & u_s''
	\end{pmatrix}.
\end{equation}
\subsection{Roots of Green's function}
We give here the general expressions of the four poles  $(\pm R q_{\pm})$ of the Green's function:
\begin{widetext}
	\begin{equation} \label{eq:Roots}
	R^2 q_{\pm}^2 = \frac{ Y(1-X^2)-X }{ Y(1-X^2)-1 } \pm \frac{\sqrt{\qty(Y(1-X^2)-X)^2-\qty(Y(1-X^2)-1)^2}}{Y(1-X^2)-1}.
	\end{equation}
\end{widetext}
\subsection{Green's function on line $\mathcal{A}/\mathcal{B}$ and $\mathcal{B}/\mathcal{C}$  }
The line separating the regions $\mathcal{A}$ and $\mathcal{B}$, of equation $Y=1/(2(1-X))$,  is associated with the Green's function,

\begin{equation}
G^{\mathcal{A}/\mathcal{B} }_\perp(s)=\frac{\delta \psi Z}{2 R} (1+X)  \qty|s| e^{-\qty|s|/R}
\end{equation}
The line separating the decay oscillating shapes, in region  $\mathcal{B}$ and the  buckling region  $\mathcal{C}$, is associated with an unphysical long-range Green's function written as,
\begin{equation}
G^{\mathcal{B}/\mathcal{C} }_{\perp}(s)=-\frac{\delta \psi R Z}{4}(1+X)\left(\delta(s)+\frac{\qty|s|}{2 R^2}\right)
\end{equation}
This behavior is due to the simultaneous  cancellation of the prefactor of the $q^4$ term of the Green's function denominator (effective bending constant) and of the constant term of the Green's function.  

\subsection{Derivation of the composition expression, Eq.~(\ref{compo})}

Consider the protonation of CL$^{2-}$, Eq.~(\ref{eq:chemical}), in the reference (undeformed) cylinder:
\begin{equation}
	\phi_0=\frac{[CL^{2-}]_0}{ [HCL^{-}]_0 + [CL^{2-}]_0} = \frac{R}{1-R},
\end{equation}
with $R=\frac{[CL^{2-}]_0}{ [HCL^{-}]_0}$. Using the equilibrium condition from pKa we have 
\begin{equation}
	K_{\rm a} = h_0 \frac{[CL^{2-}]_0}{ [HCL^{-}]_0} = \frac{h_0 \phi_0}{1-\phi_0}.
\end{equation}
Applying the previous condition to the deformed state (up to first order) one obtains
\begin{equation}
	\psi(s) = -(1-\phi_0)\frac{h(s)}{h_0}.
\end{equation}

\bibliographystyle{unsrt}
\bibliography{toni}

\end{document}